\documentclass[mathleft
]{an}
\usepackage{graphicx}
\usepackage{times}
\def\xmmn{{\it XMM-Newton~\/}}

\def\hst{{\it HST~\/}}
\def\chan{{\it Chandra~\/}}

\def\ergsec{{\rm ~erg~s^{-1}}}

\def\H0{{\rm ~km~s^{-1}~Mpc^{-1}}}

\def\kmsec{{\rm ~km~s^{-1}}}

\def\la{\mathrel{\hbox{\rlap{\hbox{\lower4pt\hbox{$\sim$}}}{\raise2pt\hbox{$<$}}}}}
\def\ga{\mathrel{\hbox{\rlap{\hbox{\lower4pt\hbox{$\sim$}}}{\raise2pt\hbox{$>$}}}}}

\def\d25{D$_{25}$}

\def\.25{0.25 keV\thinspace}

\begin{document}

\Pagespan{789}{}
\Yearpublication{2006}%
\Yearsubmission{2005}%
\Month{11}%
\Volume{999}%
\Issue{88}%

\title{(No) dynamical constraints on the mass of the black hole in two ULXs}

\author{T.P. Roberts\inst{1}\fnmsep\thanks{Corresponding author:
  \email{t.p.roberts@durham.ac.uk}\newline}
\and  J.C. Gladstone\inst{2}
\and  A.D. Goulding \inst{1}
\and  A.M. Swinbank \inst{1}
\and  M.J. Ward \inst{1}
\and  M.R. Goad \inst{3}
\and  A.J. Levan \inst{4}
}
\titlerunning{(No) dynamical constraints on the mass of two ULXs}
\authorrunning{T.P. Roberts et al.}
\institute{
Department of Physics, Durham University, South Road, Durham DH1 3LE, UK
\and 
Department of Physics, University of Alberta, Edmonton, Alberta, T6G
2G7, Canada
\and 
X-ray and Observational Astronomy Group, Dept. of Physics \&
Astronomy, University of Leicester, University Road, Leicester LE1
7RH, UK
\and 
Department of Physics, University of Warwick, Coventry CV4 7AL, UK }

\received{}
\accepted{}
\publonline{}


\abstract{We present the preliminary results of two Gemini campaigns to
constrain the mass of the black hole in an ultraluminous X-ray source
(ULX) via optical spectroscopy.  Pilot studies of the optical
counterparts of a number of ULXs revealed two candidates for
further detailed study, based on the presence of a broad He {\small II}
4686 \AA\, emission line.  A sequence of 10 long-slit spectra were
obtained for each object, and the velocity shift of the ULX
counterpart measured.  Although radial velocity variations are
observed, they are not sinusoidal, and no mass function is obtained.
However, the broad He {\small II} line is highly variable on timescales
shorter than a day.  If associated with the reprocessing of X-rays in
the accretion disc, its breadth implies that the disc must be close to
face-on.}

%

\maketitle

\section{Introduction}

The key uncertainty responsible for driving the study of ultraluminous
X-ray sources (ULXs) over the past decade is the mass of the compact
objects powering this extraordinary phenomenon.  This unknown has been
addressed by various ingenious methods, drawing evidence from across
the electromagnetic spectrum, many of which are discussed elsewhere in
these proceedings.  (A separate discussion of many of these methods,
and their results, is presented in Zampieri \& Roberts 2009).  A
consensus has emerged that ULXs are powered by accretion onto a black
hole; but the question of the mass of these black holes remains
unanswered.

The reason that ULX masses remain mired in controversy is that none of
the currently utilised methods provides a direct, unambiguous
measurement of the black hole mass.  This is perhaps best exemplified
by X-ray spectral analyses.  A great deal of progress has been made in
recent years on the basis of \chan and \xmmn spectra, firstly
identifying soft excesses consistent with the cool accretion disc
signature one would expect from a $\sim 1000 M_{\odot}$
intermediate-mass black hole (e.g.  Miller et al. 2003; Miller, Fabian
\& Miller 2004).  Latterly this interpretation has been strongly
challenged by the detection of a spectral break at energies of a few
keV, identified in the best quality \xmmn data for a wide range of
ULXs.  This implies that ULXs are operating in an unfamiliar spectral
state, most likely associated with super-Eddington processes
(Stobbart, Roberts \& Wilms 2006; Roberts 2007; Gladstone, Roberts \&
Done 2009; also Gladstone, these proceedings).  This `ultraluminous
state' appears to display the characteristic imprint of a strong
outflowing wind, as predicted for super-Eddington emission
(e.g. Begelman, King \& Pringle 2006; Poutanen et al. 2007), and so
implies that ULXs harbour small, $\la 100 M_{\odot}$ black holes.

However, neither of the above examples provides a direct mass estimate
and, worse still, for most ULX X-ray spectral data degeneracy is a
problem as the quality is sufficiently poor that neither model can be
rejected\footnote{Indeed, other spectral models may also be applied in
some cases, for example slim disc spectra (e.g. Vierdayanti et
al. 2006), or reflection-dominated spectra (Caballero-Garc{\'i}a \&
Fabian 2010), to name but two.}.  Other indirect methods suffer
similarly - the QPOs detected in the power density spectra of NGC 5408
X-1 can be used to infer the presence of an IMBH (e.g. Strohmayer \&
Mushotzky 2009; also Strohmayer these proceedings), but this assumes
both a specific type of QPOs and a sub-Eddington accretion state.
Neither may be true for this source (Middleton et al. 2010; also these
proceedings).  Similarly, different models for the optical colours and
magnitudes of various ULX counterparts lead to a range of mass
estimates (e.g. Copperwheat et al. 2006; Madhusudhan et al. 2009;
Patruno \& Zampieri 2010).  It is clear therefore that we require a
`clean' test of ULX mass, untainted by model assumptions.

The obvious way forward is to perform similar experiments to those
that have a near four-decade heritage for Galactic systems: dynamical
studies based on the co-orbital motion of the accreting compact object
and its companion, donor star (see e.g. Charles \& Coe 2006; Casares
2007).  In such experiments one commonly obtains optical (and/or
UV/IR) spectra at a series of different epochs, and measures the
semi-velocity amplitude $K$ and period $P$ for the sinusoidal orbital
motions, as traced out by shifts in the observed emission and/or
absorption line wavelengths.  Measurements are usually taken from
periods when the donor star dominates the optical light, and used to
infer a mass function $f(M)$ that places a lower limit on the black
hole mass, $M_{\rm X}$.  However, it is also possible to use emission
features originating in the accretion disc to produce a mass function
(e.g. Orosz et al. 1994; Soria et al. 1998) such that

\begin{equation}
f(M) = {{M_{\rm C}^3 \sin^3 i}\over{(M_{\rm C}+M_{\rm X})^2}} = {{PK_{\rm X}^3}\over{2\pi G}},
\label{massfn}
\end{equation}

where $M_{\rm C}$ is the mass of the companion star, $K_{\rm X}$ the semi-velocity amplitude of the black hole, and $i$ the
inclination of the orbital plane to our line-of-sight, thus placing
limits on the mass of the black hole.

We can take encouragement from the recent reports of mass functions
for two extragalactic BHBs, M33 X-8 and IC 10 X-1 (Orosz et al. 2007;
Prestwich et al. 2007; Silverman \& Filippenko 2008).  However, most
ULXs are at least three times more distant than these objects, with
$m_{\rm V} \sim 20.5$ {\bf at best\/} (Motch, these proceedings) and
more typically $m_{\rm V} > 24$ (Roberts, Levan \& Goad 2008).
Furthermore, many are located in complex fields, where their
spectroscopic signal could be confused with neighbouring nebulosity
and/or stars.  Indeed, only a handful of ULXs might be accessible for
mass function measurements with current facilities.  Very few attempts have
been made to date, and these have been unsuccessful.  Kaaret \&
Corbel (2009) obtained 6 VLT/FORS observations of NGC 5408 X-1 over 3
days, but found no stellar absorption lines to base a mass function
on.  Pakull, Gris{\'e} \& Motch (2006) obtained multi-epoch data for
NGC 1313 X-2 and did find an interesting velocity shift ($\Delta v
\sim 380 \kmsec$) in the centroid of a broad He {\small II} line, but
were unable to constrain a mass function from subsequent follow-up
data (Gris{\'e} et al. 2009).  Hence the first mass function
measurement for a ULX remains a tantalising goal.  Here, we detail the preliminary results of a new attempt to obtain the
mass functions of two ULXs, Ho IX X-1 and NGC 1313 X-2, using the
Gemini observatory telescopes.

\section{Steps towards determining the dynamical mass of a ULX}

We have been developing a programme for the past few years, with the
sole aim of obtaining a dynamical mass measurement for a ULX.  The
programme has three main steps: (i) identify the optical counterparts
to ULXs on the basis of the most accurate available X-ray positions
from {\it Chandra\/}, and \hst imaging; (ii) obtain pilot optical
spectroscopy to investigate the presence of useful
spectral features; and (iii) undertake the radial velocity
measurements campaign.


\subsection{Pilot spectroscopy results}

\begin{figure}
\includegraphics[width=53mm,angle=90]{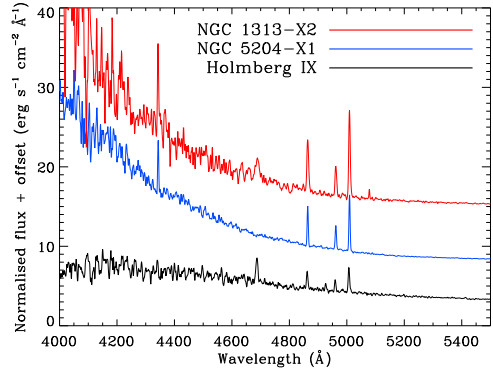}
\caption{Pilot optical long-slit spectra of the ULX counterparts.
Data were taken by the GMOS instruments on Gemini-N (NGC 5204 X-1 \&
Ho IX X-1) and Gemini-S (NGC 1313 X-2), using the B600 gratings.  The
optical magnitudes of the counterparts (extinction-corrected Vegamags,
in \hst filters) and exposure times were: $m_{555} = 23.3$, 3 hr (NGC 1313 X-2); $m_{606} = 22.3$, 0.8 hr (NGC 5204 X-1);
$m_{555} = 22.5$, 1.5 hr (Ho IX X-1).}
\label{pilotspec}
\end{figure}

In the first step we surveyed nearby ($d < 5$ Mpc) ULXs with
available \chan and \hst data, and selected relatively bright ($m_{\rm
V} \la 23.5$) and isolated objects for further study.  In
Fig.~\ref{pilotspec} we show the pilot spectra, obtained using the
GMOS instruments on the Gemini telescopes, for three X-ray luminous
($L_{\rm X} > 5 \times 10^{39} \ergsec$) ULX counterparts.  Each has
previously been identified in the literature, with positions shown by
e.g. Liu, Bregman \& Seitzer (2004, NGC 5204 X-1) and Ramsey et
al. (2006, NGC 1313 X-2 \& Ho IX X-1).

Fig.~\ref{pilotspec} shows the pilot spectra are dominated by a
relatively featureless continuum, and emission lines from the bubble
nebulae known to surround each of these three objects (Pakull \&
Mirioni 2002).  However, one interesting feature is seen in two of the
spectra: a broad He {\small II} 4686 \AA\, line is evident in the spectra
of NGC 1313 X-2 and Ho IX X-1.  It is very plausible that this line
could originate from the reprocessing of X-rays in the outer regions
of the accretion disc, so could be used to trace radial velocity
variations (as noted for NGC 1313 X-2 by Pakull et al. 2006) and
therefore place limits on the black hole mass using Eq.~\ref{massfn}.

\subsection{New Gemini spectral monitoring campaigns}

Ten follow-up Gemini observations were obtained for each of NGC 1313
X-2 (2.5 hr per observation on Gemini-S) and Ho IX X-1 (1.5 hr per
observation on Gemini-N) in semester 2009B.  Nine of the observations
of NGC 1313 X-2 were performed over a 13-day period in December 2009,
with a view to sampling over the known $\sim 6$ day photometric period
of this ULX (Liu, Bregman \& McClintock 2009).  The observations of Ho
IX X-1 were split into two blocks of 5 observations, in late December
2009 and February 2010.  Details of the data analysis will be
presented by Gladstone et al. (in prep.).  In brief, spectra were
extracted for both the counterpart, and for the surrounding nebula.
The redshift of the region containing the ULX was constrained from the
(off-ULX) nebular emission line spectrum for each observation, and
this was then used as a fiducial marker to search for relative changes
in the He {\small II} 4686 \AA\, line centroid from the
contemporaneous counterpart spectrum.

\begin{figure}
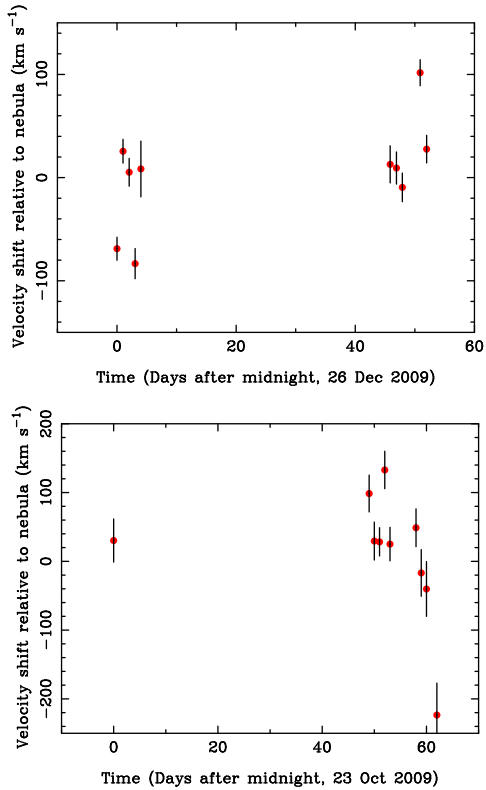

\centering
\includegraphics[width=51mm,angle=270]{rvplot_hoix.ps}\vspace*{2mm}
\includegraphics[width=51mm,angle=-90]{rvplot_1313.ps}
\caption{Observed radial velocity shifts of the ULX counterparts
relative to the bubble nebulae, measured from the broad He {\small II}
4686 \AA\, line, over the observation campaign.  {\it Top panel:\/} Ho
IX X-1.  {\it Lower panel:\/} NGC 1313 X-2.}
\label{rvs}
\end{figure}

\section{Provisional results}

\begin{figure}
\includegraphics[width=78mm]{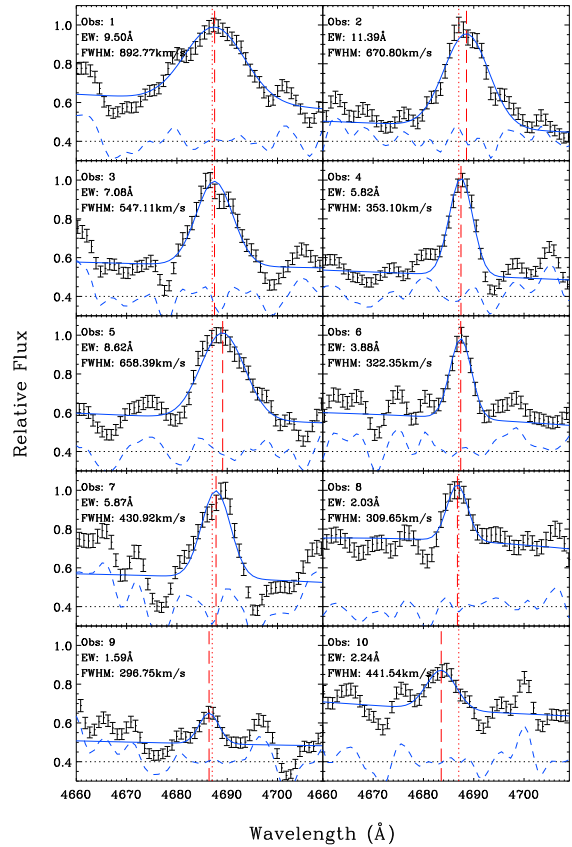}
\caption{He {\small II} 4686 \AA\, line profiles from the NGC 1313 X-2
data. The best fitting Gaussian profile is shown as a solid blue line,
and the data residuals to this fit as a dashed blue line below the
data.  The vertical dashed and dotted lines show the rest wavelength
and the line centroid, respectively.  The relative flux is shown with
the peak line flux normalised to unity for all good ($\geq 9\sigma$) line detections.  The last two detections have lower statistical significances, of 4 and $6\sigma$ respectively.}
\label{heiiplot}
\end{figure}

The results we present here are based on an initial analysis of the
data, and focus primarily on NGC 1313 X-2.  A further, more complete
analysis is in preparation by Gladstone et al. and Roberts et al.

The preliminary results show that the campaign was successful on one
count: we detected the anticipated shifts in the He {\small II}
4686 \AA\, line with respect to its local vicinity (see
Fig.~\ref{rvs}), measuring velocity shifts of $\pm 100 \kmsec$ for Ho
IX X-1, and up to $\sim 200 \kmsec$ for NGC 1313 X-2.  However, the
data was not consistent with sinusoidal variations in either case.
Simple sinusoid fitting to both datasets resulted in poor fits, with
$P = 1.73$ days, $K = 77 \kmsec$ and $\chi^2_{\nu} \sim 5$ for Ho IX
X-1; and $P = 3.01$ days, $K = 61 \kmsec$ and $\chi^2_{\nu} \sim 4$ for
NGC 1313 X-2.  Given the lack of evidence for sinusoidal
(i.e. orbital) variations in the data, we must therefore conclude that
the data does not provide a new, solid constraint on the mass function
for either object.

\begin{figure*}
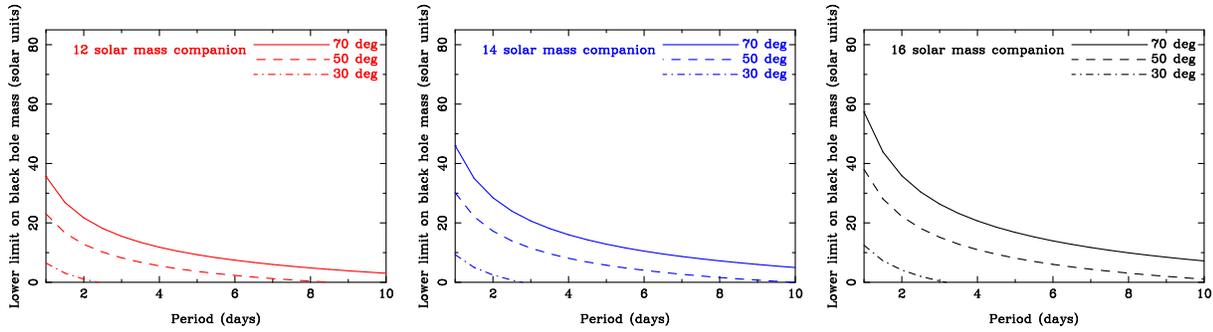

\centering
\includegraphics[width=43mm,angle=270]{bhmassplot_1313.ps}\hspace*{2mm}
\includegraphics[width=43mm,angle=270]{bhmassplot2_1313.ps}\hspace*{2mm}
\includegraphics[width=43mm,angle=270]{bhmassplot3_1313.ps}
\caption{Mass limits on the black hole in NGC 1313 X-2, assuming the
He {\small II} 4686 \AA\, line originates in the accretion disc.  We use
the $2\sigma$ upper limit of $182 \kmsec$ derived from the rms scatter of the
velocity shift, companion stellar mass estimates based on Patruno \&
Zampieri (2010), and calculate lower limits on the mass (for a range
of inclinations, as per the legend) based on Eq.~\ref{massfn}.}
\label{bhmasses}
\end{figure*}

Despite the lack of mass function measurements, the data does reveal
interesting behaviour from the ULX counterpart.  One notable feature
is the highly variable nature of the broad He {\small II} 4686 \AA\,
line.  In Fig.~\ref{heiiplot} we show the profile of this line in the
10 separate observations of NGC 1313 X-2.  Over these observations,
its FWHM varies from a peak of $\sim 900 \kmsec$ down to the
instrumental resolution ($< 290 \kmsec$), with variations of factor 2
seen in 24 hours.  This dramatic variability implies at the minimum
that the broad He {\small II} emission must be originating within 24
light-hours of the ULX.

Such variations might come from X-ray reprocessing in the outer
regions of the accretion disc.  If so, the broad lines tell us that
the disc cannot be perfectly face-on (or else we would see no line-of-sight velocity variations).  This means that there should be some
information on the black hole orbit hidden within the radial velocity
data: we therefore use the $2\sigma$ upper limit derived from the scatter in
radial velocity measurements for NGC 1313 X-2 to derive lower limits
on its black hole mass for a range of orbital periods in
Fig~\ref{bhmasses}.  However, as Kaaret \& Corbel pointed out for NGC
5408 X-1, such small measured velocities are inconsistent with the expected velocities of material in the accretion disc ($\gg 1000 \kmsec$).  A possible solution is that only a small
component of the accretion disc rotational velocity is in the
line-of-sight, i.e. the disc is very close to face-on ($i \sim
0^{\circ}$).  We see in Fig~\ref{bhmasses} that even at $i =
30^{\circ}$ there is little or no lower limit on the black hole mass,
other than at very short periods\footnote{Although removing the last two, low significance radial velocity points, significantly reduces the rms scatter to $44$ km s$^{-1}$, and this can lead to some interesting constraints, e.g. for a 16 $M_{\odot}$ companion and $30^{\circ}$ inclination, $M_{\rm BH} > 10 M_{\odot}$ over all periods up to 10 days.}.  Hence if the disc is close to face-on, little can be said
about the black hole mass.

\section{Conclusion}

The initial analyses of our campaigns to determine mass functions for
two ULXs betray no strong evidence for periodic velocity variations
from either object; and so no mass function is forthcoming.  We do see
some interesting phenomenology in the ULX counterparts, including
strong variability in the He {\small II} 4686 \AA\, line indicating it
originates very close to the ULX.  The FWHM of the broad line
indicates that the ULX may be close to face-on; interestingly,
both objects are X-ray luminous ($L_{\rm X} > 5 \times 10^{39}
\ergsec$), and models of super-Eddington discs predict that their
observed flux will be highest close to this line-of-sight
(Mineshige, these proceedings).  This is perhaps more
circumstantial evidence for models of ULXs as super-Eddington
stellar-mass systems rather than IMBHs.  In the meantime, we still
await the first determination of the mass function for a ULX.



\acknowledgements

The authors thank Manfred Pakull and Peter Jonker for suggestions on
improvements to the analysis presented in this paper, the anonymous referee for useful comments, and the
organisers of the ``Ultraluminous X-ray sources and middle-weight
black holes'' workshop for a very stimulating meeting.  This work is
based on observations obtained at the Gemini Observatory.



\end{document}